\documentclass[12pt]{article}
\begin{document}

\begin{center}
{\large Superfluid vs Ferromagnetic Behaviour in a Bose Gas of Spin-1/2 
Atoms}

\

S. Ashhab

{\footnotesize Department of Physics, Ohio State University, 174 West 
Eighteenth Avenue, Columbus, Ohio 43210}

August 31, 2004

\vspace{2cm}

{\bf Abstract}
\end{center}
We study the thermodynamic phases of a gas of spin-1/2 atoms in the 
Hartree-Fock approximation. Our main result is that, for repulsive or 
weakly-attractive inter-component interaction strength, the superfluid and 
ferromagnetic phase transitions occur at the same temperature. For 
strongly-attractive inter-component interaction strength, however, the 
ferromagnetic phase transition occurs at a higher temperature than the 
superfluid phase transition. We also find that the presence of a 
condensate acts as an effective magnetic field that polarizes the normal 
cloud. We finally comment on the validity of the Hartree-Fock 
approximation in describing different phenomena in this system.

\newpage

In recent years, studies of multi-component Bose-Einstein Condensates 
(BEC) have revealed a variety of interesting phenomena that reflect 
qualitatively different ideas from spinless condensates 
\cite{Myatt,Stenger,Ohmi,Ho1,Law,Leanhardt,Pu,Schmaljohann,Chapman}. Gases 
of spin-1/2 \cite{Myatt}, spin-1 
\cite{Stenger,Ohmi,Ho1,Law,Leanhardt,Pu,Schmaljohann} and spin-2 
\cite{Schmaljohann,Chapman} atoms have been studied in considerable 
detail, both experimentally and theoretically. New physical phenomena 
include fragmented states \cite{Law}, coreless vortices \cite{Leanhardt} 
and complex spin dynamics \cite{Pu,Schmaljohann,Chapman}. Truely 
spin-1/2 particles are fermions. However, a gas of bosonic atoms each of 
which can occupy two different internal states can be treated as a gas of 
spin-1/2 atoms that obey Bose statistics (note that in nonrelativistic 
systems, there is no {\it a priori} connection between spin and 
statistics). Examples of such gases are mixtures of $^{87}$Rb atomic gases 
in two hyperfine states and spin-polarized Hydrogen \cite{Myatt,Bonalde}.

It is well known that in a gas of noninteracting spin-1/2 bosonic atoms, 
the emergence of superfluid order is accompanied by the emergence of 
ferromagnetic order \cite{Siggia,Ho2}. A natural question that arises in 
that context is whether these two orders would remain connected 
if we include the effects of interatomic interactions. Perhaps a clearer 
way to pose the question is as follows: is it possible to have one of 
those two orders (i.e. superfluid or ferromagnetic) without having the 
other? One can give a number of elementary arguments that hint one way or 
the other. On the one hand, in the noninteracting system the connection 
between the two orders follows from the symmetry of the wave function and 
is not related to any thermodynamic arguments (except for saying that the 
normal cloud is unpolarized). That would suggest a rather robust 
connection between superfluid and ferromagnetic behaviours. On the other 
hand, if one considers the form of the order parameters or the symmetry of 
the Hamiltonian, there is no reason to believe that the two orders must be 
related. Keeping in mind certain caveats, the superfluid order parameter 
can be chosen as $\left|\langle \hat{\psi}_{\uparrow}({\bf r}) \rangle 
\right|^2 + \left|\langle \hat{\psi}_{\downarrow}({\bf r}) \rangle 
\right|^2$, whereas the ferromagnetic order parameter can be 
chosen as $\sum_{ij} \langle \hat{\psi}_i^{\dagger}({\bf r}) 
\vec{\sigma}_{ij} \hat{\psi}_j({\bf r}) \rangle$, where 
$\vec{\sigma}_{ij}$ is the vector of Pauli spin matrices, and the creation 
and annihilation operators used above will be explained when we give the 
Hamiltonian of the system below. The noninteracting Hamiltonian obeys 
SU(2) symmetry, whereas the interacting Hamiltonian obeys 
U(1)$\otimes$U(1) symmetry. Furthermore, it has been shown recently by 
Yang and Li that the gound state of this system with no spin-dependent 
interactions is ferromagnetic, even if the interactions are strong enough 
to completely suppress superfluidity \cite{Yang}. The energies of 
low-lying excitations, however, also drops to zero for such strong 
interactions. It is therefore not obvious, at first sight, whether or not 
one order can exist without the other at finite temperature. A similar 
problem for a Bose gas of spin-1 atoms with ferromagnetic interactions has 
been considered by Gu and Klemm \cite{Gu}. One clear difference between 
spin-1/2 and spin-1 atoms is that with the latter one has a parameter that 
determines whether interatomic interactions are ferromagnetic or 
antiferromagnetic, even in a rotation-symmetric system. With spin-1/2 
atoms, apart from some obvious cases that we shall explain shortly, it is 
in general not immediately clear whether a certain set of interaction 
parameters describes ferromagnetic interatomic interactions. In fact, as 
explained above, a rotation-symmetric gas of spin-1/2 atoms always 
exhibits ferromagnetic behaviour.

In the present Paper, we shall try to obtain the answer to the above-posed 
question in the Hartree-Fock (i.e. mean-field) approximation. We do that 
by calculating the free energy as a function of the macroscopic 
thermodynamic variables and minimizing the free energy with respect to 
those variables. Since similar approximations, as well as a number of 
systematic field-theoretic calculations, are not reliable in predicting 
the superfluid transition temperature in a spinless system (see e.g. 
\cite{Baym,Fetter}, also see Appendix A), our results are necessarily 
plagued by the same type of unreliability, and they cannot be considered 
conclusive. The advantage of using the the Hartree-Fock approximation, 
however, is that whenever it gives correct results, it gives them with a 
simple physical explanation. At the end of our treatment, we shall give 
our intuitive assessment of which results we believe describe real 
physical phenomena and which results we believe are merely artifacts of 
the Hartree-Fock approximation.

One rather trivial phenomenon in the context of spin-1/2 atoms occurs 
when the two internal states have different internal energies. In that 
case, the gas can be almost completely polarized at microkelvin 
temperatrues, assuming chemical equilibrium between the two spin states is 
reached. Although that phenomenon can be considered one form of 
ferromagnetic behaviour, we are not interested in it. Another 
ferromagnetic behaviour that we do not wish to consider here occurs when 
the two intra-component interaction strengths are sufficiently different. 
In that case it can, under certain conditions, be favourable for the 
majority of the atoms to occupy a single spin state. The system we shall 
consider in this Paper, which is also the most experimentally-relevant 
system of spin-1/2 atoms, is the one where the total spin along some axis 
is conserved, but the total spin perpendicular to that axis is not 
conserved. In the language of spin systems, one says that the longitudinal 
spin-relaxation time $T_1$ is much longer than the time scale of 
performing the experiment, whereas the transverse spin-relaxation time 
$T_2$ is shorter than the time scale of performing the experiment. We take 
the longitudinal axis to be the $z$-axis. Note that in the situation 
described above, and assuming that one starts with no net magnetization 
along the $z$-axis, the macroscopic magnetization expected for a 
condensate of noninteracting atoms lies entirely in the $xy$-plane.

Let us take a Bose gas of spin-1/2 atoms in a three dimensional box. The 
Hamiltonian of the system can be expressed as:

\noindent
\begin{equation}
\hat{H} = \hat{H}_{\rm kin} + \hat{H}_{\rm int},
\end{equation}

\noindent where

\noindent
\begin{eqnarray}
\hat{H}_{\rm kin} & = & -\frac{\hbar^2}{2m} \int d{\bf r} 
\sum_{\sigma=\uparrow,\downarrow} \hat{\psi}^{\dagger}_{\sigma}({\bf r})
\nabla^2 \hat{\psi}_{\sigma}({\bf r}),
\\
\hat{H}_{\rm int} & = & \int d{\bf r} 
\sum_{\sigma,\sigma'=\uparrow,\downarrow} \frac{g_{\sigma\sigma'}}{2} 
\hat{\psi}^{\dagger}_{\sigma}({\bf r}) \hat{\psi}^{\dagger}_{\sigma'}({\bf 
r}) \hat{\psi}_{\sigma'}({\bf r}) \hat{\psi}_{\sigma}({\bf r}),
\end{eqnarray}

\noindent
$m$ is the atomic mass, $\hat{\psi}^{\dagger}_{\sigma}({\bf r})$ is an 
operator that creates an atom at position ${\bf r}$ in spin state 
$\sigma$ (the spin states $\uparrow$ and $\downarrow$ correspond to the 
$z$-component of atomic spin being equal to $\pm 1/2$, respectively), 
$\hat{\psi}_{\sigma}({\bf r})$ is its Hermitian 
conjugate, $g_{\sigma\sigma'}=4\pi\hbar^2a_{\sigma\sigma'}/m$, and 
$a_{\sigma\sigma'}$ is the scattering length between two atoms in spin 
states $\sigma$ and $\sigma'$. We assume that the number of atoms in 
the $\sigma$ and $\sigma'$ spin states is individually conserved, whereas 
the total spin in the $xy$-plane is not conserved. Furthermore, we shall 
assume that there is an equal number of atoms in the two spin states 
$\uparrow$ and $\downarrow$ \cite{EqualNumbers}. In order to avoid dealing 
with the possibility of mechanical collapse of the gas or phase separation 
of the two components, we shall assume that both $g_{\uparrow\uparrow}$ 
and $g_{\downarrow\downarrow}$ are positive and that 
$g_{\uparrow\downarrow}^2 < g_{\uparrow\uparrow}g_{\downarrow\downarrow}$.

We now derive an expression for the free energy of the above-described
system as a function of the macroscopic thermodynamic variables and use it 
to determine the state of the system for a given temperature. The 
macroscopic variables are the number of atoms in the condensate $N_o$, the 
total spin of the condensate ${\bf S}_C$, the number of atoms in the 
normal cloud $N_N$, and the total spin of the normal cloud ${\bf S}_N$. 
We use the canonical ensemble, where the constraint of fixed total number 
of atoms $N_o+N_N$ is imposed explicitly. Since we are dealing with a 
macroscopic system, we treat the above variables as classical 
variables. A condensate of spin-1/2 atoms is ferromagnetic as a result of 
Bose symmetry \cite{Siggia,Ho2}. Its total spin $S_C$ is equal to $N_o/2$ 
[i.e. $\hat{S}_C^2=N_o/2(N_o/2+1)$]. On the other hand, the only 
constraint on the total spin of the normal cloud is that $|{\bf S}_N|\leq 
N_N/2$. Therefore, we express the $x$, $y$ and $z$ components of the total 
spin of the condensate as $(N_o/2 \sin\theta_o\cos\varphi_o,N_o/2 
\sin\theta_o\sin\varphi_o,N_o/2 \cos\theta_o)$ and those of the normal 
cloud as $(N_N s_{N\perp}\cos\varphi_N,N_N s_{N\perp}\sin\varphi_N,-N_o/2 
\cos\theta_o)$. Note that with the above values of condensate and 
normal-cloud spins, we have taken into account our assumption that the 
$z$-component of the total spin must vanish. For any given set of values 
of the thermodynamic variables, the free energy is given by:

\noindent
\begin{equation}
F = \langle \hat{H}_{\rm kin} \rangle + \langle \hat{H}_{\rm int} \rangle 
- TS
\end{equation}

\noindent
where $T$ is the temperature and $S$ is the entropy of the system. The 
entropy $S$ and the expectation values are calculated by considering all 
the different microscopic configurations corresponding to the given 
values of the macroscopic variables. We first consider the first 
and third terms of the free energy, which we denote by $F_{\rm ideal}$. 
The condensate does not contribute to those terms in the free energy. 
Therefore we only need to evaluate $F_{\rm ideal}$ for a normal cloud of 
$N_N$ atoms with total spin $S_N$. As we shall see below, the interaction 
energy is constant to leading order for all the different microscopic 
configurations in the Hartree-Fock thermodynamic ensemble for given $N_N$ 
and $S_N$. Therefore, $F_{\rm ideal}$ is given by the same expression that 
it takes in the noninteracting system. The free energy of a spinless 
noninteracting uniform Bose gas is given by \cite{Huang}:

\noindent
\begin{eqnarray}
F_{\rm ideal} & \equiv & \langle \hat{H}_{\rm kin} \rangle -TS
\\
& = & N_N k_B T \left( \ln z - \frac{g_{5/2}(z)}{g_{3/2}(z)} \right),
\end{eqnarray}

\noindent
where $k_B$ is Boltzmann's constant and $z$ is given by

\noindent
\begin{eqnarray}
g_{3/2}(z) & = & n_N \lambda_t^3,
\\
g_{j}(z) & \equiv & \sum_{l=1}^{\infty} \frac{z^l}{l^j},
\\
\lambda_t & \equiv & \sqrt{\frac{2\pi\hbar^2}{mk_BT}},
\end{eqnarray}

\noindent
$n_N=N_N/V$, and $V$ is the volume of the sample. $F_{\rm ideal}$ of the 
system at hand is given by the sum of two terms of the above form of 
$F_{\rm ideal}$ for two independent Bose gases, one with $N_N(1/2+s_N)$ 
atoms and the other with $N_N(1/2-s_N)$ atoms, where 
$s_N=\sqrt{s_{N\perp}^2 + N_o^2 \cos^2\theta_o/4N_N^2}$:

\noindent
\begin{equation}
\frac{F_{\rm ideal}}{N_N k_B T} = \left( \frac{1}{2} + s_N \right) 
\left( \ln z_+ - \frac{g_{5/2}(z_+)}{g_{3/2}(z_+)} \right) + \left( 
\frac{1}{2} - s_N \right) \left( \ln z_- - 
\frac{g_{5/2}(z_-)}{g_{3/2}(z_-)} \right),
\end{equation}

\noindent where $z_{\pm}$ are given by

\noindent
\begin{equation}
g_{3/2}(z_{\pm}) = n_N \lambda_t^3 \left( \frac{1}{2} \pm s_N \right).
\label{eq:Define_z}
\end{equation}

\noindent
Note that since $z_+$ cannot exceed the value 1, $s_N$ must obey the 
condition $n_N\lambda_t^3 (1/2+s_N)\leq g_{3/2}(1)$. Also note that 
$F_{\rm ideal}$ is a monotonically increasing function of $s_N$. If we 
take the limit $s_N\rightarrow 0$, we find that:

\noindent
\begin{equation}
\frac{F_{\rm ideal}}{N_Nk_BT} = \alpha + \gamma s_N^2 + \eta s_N^4 + 
O(s_N^6),
\label{eq:F_expansion}
\end{equation}

\noindent where

\noindent
\begin{eqnarray}
\alpha & = & \ln z - \frac{g_{5/2}(z)}{g_{3/2}(z)},
\\
\gamma & = & 2 \frac{g_{3/2}(z)}{g_{1/2}(z)},
\\
\eta & = & \frac{2g_{3/2}^3(z)}{3} \left( 
\frac{3g_{-1/2}^2(z)}{g_{1/2}^5(z)} - \frac{g_{-3/2}(z)}{g_{1/2}^4(z)} 
\right),
\end{eqnarray}

\noindent
and $z$ is evaluated from Eq. (\ref{eq:Define_z}) with $s_N=0$. When 
$z=0$, $\gamma=2$ and $\eta=4/3$, and as $z\rightarrow 1$, 
$\gamma\propto {\rm const.} \sqrt{1-z}$ whereas $\eta$ decreses slightly 
from the value $4/3$ and remains finite. We now calculate the interaction 
energy in the Hartree-Fock approximation. In that approximation we assume 
that there is no coherence between states of different relative momentum 
of a pair of interacting atoms (i.e. $\langle a_{k_1\sigma_1}^{\dagger} 
a_{k_2\sigma_2}^{\dagger} a_{k_3\sigma_3} a_{k_4\sigma_4} \rangle$ 
vanishes unless the momenta of the creation operators match those of the 
annihilation operators, not just the sum of the momenta). The two-particle 
correlation functions needed to evaluate the interaction energy can then 
be straightforwardly calculated to give:

\noindent
\begin{eqnarray}
\langle \hat{\psi}^{\dagger}_{\uparrow}({\bf r}) 
\hat{\psi}^{\dagger}_{\uparrow}({\bf r}) \hat{\psi}_{\uparrow}({\bf r}) 
\hat{\psi}_{\uparrow}({\bf r}) \rangle & = & n_o^2 
\cos^4\frac{\theta_o}{2} + \frac{n_N^2}{2} \left( 1 - \frac{n_o}{n_N} 
\cos\theta_o \right)^2 + 2 n_o n_N \cos^2\frac{\theta_o}{2} \left( 1 - 
\frac{n_o}{n_N} \cos\theta_o \right)
\\
\langle \hat{\psi}^{\dagger}_{\downarrow}({\bf r})
\hat{\psi}^{\dagger}_{\downarrow}({\bf r}) \hat{\psi}_{\downarrow}({\bf 
r}) \hat{\psi}_{\downarrow}({\bf r}) \rangle & = & n_o^2 
\sin^4\frac{\theta_o}{2} + \frac{n_N^2}{2} \left( 1 + \frac{n_o}{n_N} 
\cos\theta_o \right)^2 + 2 n_o n_N \sin^2\frac{\theta_o}{2} \left( 1 + 
\frac{n_o}{n_N} \cos\theta_o \right)
\\
\langle \hat{\psi}^{\dagger}_{\uparrow}({\bf r})
\hat{\psi}^{\dagger}_{\downarrow}({\bf r}) \hat{\psi}_{\downarrow}({\bf
r}) \hat{\psi}_{\uparrow}({\bf r}) \rangle & = & \left( \frac{n_o+n_N}{2} 
\right)^2 + n_N^2 s_{N\perp}^2 + n_o n_N s_{N\perp} \sin\theta_o 
\cos(\varphi_o-\varphi_N)
\end{eqnarray}

\noindent where $n_o=N_o/V$. By direct substitution of the above 
correlation functions, one can find the expression for the interaction 
energy $\langle \hat{H}_{\rm int} \rangle$ of the system. The state 
of the system can now be determined by minimizing the free energy:

\noindent
\begin{eqnarray}
\frac{F}{V} & = & f_o - g_- n_o^2 \cos\theta_o - \frac{g_+}{2} n_o^2 
\cos^2\theta_o + g_{\uparrow\downarrow} n_N^2 s_{N\perp}^2 + 
g_{\uparrow\downarrow} n_o n_N s_{N\perp} \sin\theta_o 
\cos(\varphi_o-\varphi_N) 
\nonumber
\\
& & + \frac{F_{\rm ideal}(N_N,k_BT,s_{N\perp},\cos\theta_o)}{V},
\label{eq:Free_Energy}
\end{eqnarray}

\noindent where

\noindent
\begin{eqnarray}
f_o & = & \frac{\bar{g}}{2} n_o^2 + \tilde{g} n_N^2 + 2 \tilde{g} n_o n_N
\nonumber
\\
& = & \tilde{g} (n_o+n_N)^2 - \frac{g_+}{2} n_o^2
\\
\bar{g} & = & \frac{g_{\uparrow\uparrow} + g_{\downarrow\downarrow} + 
2 g_{\uparrow\downarrow}}{4}
\\
\tilde{g} & = & \frac{g_{\uparrow\uparrow} + g_{\downarrow\downarrow} +
g_{\uparrow\downarrow}}{4}
\\
g_{\pm} & = & \frac{g_{\uparrow\uparrow} \pm g_{\downarrow\downarrow}}{4}
\end{eqnarray}

\noindent
Note that the fourth and fifth terms in Eq. (\ref{eq:Free_Energy}), among 
others, result from exhange-interactions. Those two terms describe 
scattering processes where an atom with spin $\uparrow$ and another with 
spin $\downarrow$ exchange their momenta. In the expression for the free 
energy (Eq. \ref{eq:Free_Energy}), the second and third terms favour 
maximizing $|\cos\theta_o|$ (i.e. having finite values of the 
$z$-component of the condensate and normal cloud), whereas the fifth and 
sixth terms favour taking $\cos\theta_o=0$. The second term can be 
understood quite intuitively as follows. Since the interaction term 
is magnified in the normal cloud as compared to the condensate (because of 
the exchange terms), the normal atoms tend to accumulate in the spin state 
with less-repulsive interactions. For example, if we take 
$g_{\uparrow\uparrow} > g_{\downarrow\downarrow}$, we find that the normal 
cloud will have an excess of atoms in the $\downarrow$ spin state, leaving 
the condensate with an excess of atoms in the $\uparrow$ spin state. Since 
in this Paper we are not interested in that phenomenon, we eliminate it by 
taking $g_{\uparrow\uparrow} = g_{\downarrow\downarrow}$. The third term 
in Eq. (\ref{eq:Free_Energy}) can also be understood by considering the 
absence of exchange interaction terms within the condensate. Since we are 
requiring that half the atoms have spin $\uparrow$ and the other half have 
spin $\downarrow$, it is straightforward to see that in order to gain the
greatest reduction in interaction energy, the condensate atoms tend to
accumulate in the same spin state (either $\uparrow$ or $\downarrow$). The 
competition between the third term and last two terms of Eq. 
(\ref{eq:Free_Energy}) determines whether the condensate and normal cloud 
will have any net magnetization along the $z$-axis. Clearly, at low enough 
temperatures, the third term will win, and one will have a finite 
$z$-component of the polarization of the condensate and normal cloud (note 
that the net polarization of the entire cloud must vanish, which is one 
of our main assumptions). That result is quite interesting in its own 
right. However, since the main question addressed in this Paper concerns 
the phase transitions, we focus our attention on those relatively-high 
temperatures. We minimize the free energy with respect to $n_o$, 
$\cos\theta$ and $s_{N\perp}$ at any given combination of the parameters 
$g_{\uparrow\uparrow}$, $g_{\uparrow\downarrow}$ and $T$ to find the 
thermodynamic phases of the system. The results of a numerical calculation 
of the order of the phase transition are shown in Fig. 1. It is also worth 
making some analytical remarks about the bahaviour of the system. For 
clarity we address the following two cases separately (in the following 
remarks, we shall implicitly use a result that we found from our numerical 
calculation, namely that $cos\theta$ remains negligibly small close to the 
transition temperatures):
\newline
{\it Case 1}: $g_{\uparrow\downarrow}>0$. If $n_o\neq 0$, the free energy 
is minimized by taking $\varphi_o-\varphi_N=\pi$ (i.e the polarizations 
of the condensate and normal cloud point in opposite directions) and a 
finite value of $s_{N\perp}$. In the limit $n_o/n_N\ll 1$,

\noindent
\begin{equation}
s_{N\perp} = \frac{g_{\uparrow\downarrow}n_o}{2(\gamma k_BT 
+ g_{\uparrow\downarrow}n_N)} + 
O\left(\frac{k_BTn_o^3}{(\gamma k_BT+g_{\uparrow\downarrow}n_N)^4}\right)
\end{equation}

\noindent
Note that above the BEC critical temperature $T_c$, i.e. when $n_o=0$, the 
normal cloud is not polarized at all. Below $T_c$ the physics can be 
understood in terms of the condensate acting as an effective magnetic 
field that partially polarizes a paramagnetic normal cloud. Since the 
superfluid transition occurs at a temperature where $\gamma k_BT\sim 
g_{\uparrow\uparrow} n$ (see Appendix A), we find that: 

\noindent
\begin{equation}
n_N s_{N\perp} < \frac{n_o}{2}
\end{equation}

\noindent
Therefore, the polarization of the normal cloud is smaller than that of 
the condensate, and the net polarization does not vanish. Note that if we 
calculate the exact expression for $n_N s_{N\perp}$, substitute it in 
Eq. (\ref{eq:Free_Energy}) and minimize $F$ with respect to $n_o$ (keeping 
$n_o+n_N$ fixed), we find that the global minimum of $F$ jumps 
discontinuously from a point with $n_o=0$ to a point with a finite value 
of $n_o$. That would suggest a first-order phase transition to the 
superfluid phase, which is also ferromagnetic. However, the Hartree-Fock 
approximation predicts a first-order phase transition in a spinless Bose 
gas \cite{Baym} (also see Appendix A), and we therefore suspect that this 
result must be an artifact of the approximation.
\newline
{\it Case 2}: $g_{\uparrow\downarrow}<0$. In this case the free energy is 
minimized by taking $\varphi_o-\varphi_N=0$ (i.e the polarizations of the 
condensate and normal cloud point in the same direction). The term 
$g_{\uparrow\downarrow}n_N^2s_{N\perp}^2$ favours a polarized normal 
cloud. That suggests that the gas might exhibit ferromagnetic behaviour 
even if $n_o=0$, as we shall see shortly. Using the small $s_{N\perp}$ 
expression for the ideal-gas free energy (Eq. \ref{eq:F_expansion}) we 
find that the free energy is minimized by choosing

\noindent
\begin{equation}
s_{N\perp} = \frac{|g_{\uparrow\downarrow}|n_o}{2(\gamma k_BT
-|g_{\uparrow\downarrow}|n_N)} + O \left( \frac{k_BT n_o^3}{(\gamma 
k_BT-|g_{\uparrow\downarrow}|n_N)^4} \right),
\end{equation}

\noindent
if $\gamma k_BT-|g_{\uparrow\downarrow}|n_N \gg 
|g_{\uparrow\downarrow}|n_o$. As in {\it case 1}, the condensate acts as 
an effective magnetic field that polarizes the normal cloud. One can also 
immediately see that the normal gas exhibits ferromagnetic behaviour when 
$\gamma k_BT-|g_{\uparrow\downarrow}|n_N$ becomes negative. Note that 
since $\gamma=0$ when $T/T_c^o\leq 1$, the ferromagnetic phase transition 
must occur at a temperature higher than that of the ideal-gas BEC phase 
transition $T_c^o$. Assuming that $n_o=0$ just below the ferromagnetic 
phase transition, we find that that transition occurs at a temperature 
where:

\noindent
\begin{equation}
2 \frac{g_{3/2}(z)}{g_{1/2}(z)} k_BT_{\rm ferro} = 
|g_{\uparrow\downarrow}| n_N,
\label{eq:ferro_criterion}
\end{equation}

\noindent
where $z$ is given by:

\noindent
\begin{equation}
\frac{g_{3/2}(z)}{g_{3/2}(1)} = \left( \frac{T_c^o}{T_{\rm ferro}} 
\right)^{3/2}.
\end{equation}

\noindent
Just below the transition temperature, the spin grows as:

\noindent
\begin{equation}
s_N = \sqrt{\frac{\gamma k_B (T_{\rm ferro}-T)}{2 \eta}}.
\end{equation}

\noindent
Note that if $|g_{\uparrow\downarrow}|$ is smaller than a certain value, 
the ferromagnetic transition temperature can be smaller than the 
superfluid transition temperature, which is also higher than $T_c^o$ for 
positive $g_{\uparrow\uparrow}$. In that case there would be a single 
(first-order) phase transition, just as in the case 
$g_{\uparrow\downarrow}>0$. However, as can be seen from Fig. 1 and as 
explained in Appendices A and B, there is a region in parameter space 
where the ferromagnetic transition temperature is higher than the 
superfluid transition temperature.

We now summarize our results and comment on them. We found that when 
$g_{\uparrow\downarrow}>0$, there is a single transition to a 
phase that has both superfluid and ferromagnetic order, and the phase 
transition is first order. We believe that the first-order nature of the 
phase transition is an artifact of the Hartree-Fock approximation. 
However, since the physical mechanisms favouring an unpolarized 
gas (interaction energy) and those favouring an unpolarized normal gas 
(entropy) are both real physical mechanisms, we suspect that as soon as 
there is a finite fraction of atoms in the condensate, the induced 
polarization of the normal cloud will be smaller than the polarization 
of the condensate, and there will in fact be a single phase transition to 
a state with both superfluid and ferromagnetic order. In the case 
$g_{\uparrow\downarrow}<0$, we found that it is possible, for large 
enough $|g_{\uparrow\downarrow}|$, to have two phase transitions. At 
temperature $T_{\rm ferro}$, we found a second-order phase transition to a 
ferromagnetic phase with no superfluid order. We suspect that that result 
will persist even beyond the Hartree-Fock approximation, because that 
transition occurs above the superfluid transition temperature, and 
therefore fluctuations in the superfluid order parameter are expected to 
be negligible. At a lower termperature we found a first-order phase 
transition to a superfluid phase (with ferromagnetic order). As above, we 
believe that the superfluid phase transition will also be second-order. 
Based on similar elementary arguments alone, we cannot comment on the 
location of the boundary separating the single-transition and 
double-transition regimes.

In conclusion, we have performed a Hartree-Fock (mean-field) calculation 
to study the phase transitions in a gas of spin-1/2 bosonic atoms. We 
found that it is possible to have ferromagnetic order with no superfluid 
order, but not vice versa. We also found that the phase transition to the 
ferromagnetic non-superfluid phase is second-order, whereas any transition 
to a superfluid phase is first order. We suspect that the result of a 
nonsuperfluid ferromagnetic phase describes a real physical phenomenon. 
Judging from the Hartree-Fock results in the spinless case, however, we 
believe that in the real system, all phase transitions will turn out to be 
second order in nature. Due to the unreliability of the Hartree-Fock 
approximation near the superfluid transition temperature, further study is 
required to confirm or refute the results of this Paper.

The author would like to thank T.-L. Ho for useful discussions and Q. Gu 
for useful correspondence about the results of Ref. \cite{Gu}. This work 
was supported by the National Science Foundation through Grant Nos. 
DMR-0071630 and DMR-0109255 and by NASA through Grant Nos. NAG8-1441 and 
NAG8-1765.

\noindent {\bf Appendix A:}

In this Appendix we show that the Hartree-Fock approximation predicts a 
first-order phase transition in a weakly-interacting spinless 
Bose gas \cite{Baym} (Here we are assuming repulsive interactions). Using 
similar arguments to those used in the main text, we find that the free 
energy density as a function of condensate number density $n_o$ and normal 
cloud number density $n_N$ is given by:

\noindent
\begin{equation}
\frac{F}{V} = n_N \left( \ln z - \frac{g_{5/2}(z)}{g_{3/2}(z)} \right) - 
\frac{g}{2} n_o^2 + g (n_o+n_N)^2,
\end{equation}

\noindent where

\noindent
\begin{equation}
g_{3/2}(z) = n_N \left(\frac{2\pi\hbar^2}{mk_BT}\right)^{3/2},
\label{eq:g_App}
\end{equation}

\noindent
with the constraint that the right-hand side of Eq. (\ref{eq:g_App}) 
cannot be greater than $g_{3/2}(1)$. The dependence of $(F(n_o)-F(0))/V$ 
on $n_o$ (keeping the total density fixed) is shown schematically in Fig. 
A1. Clearly this calculation predicts a first-order phase transition at a 
temperature $T_c>T_c^o$, where $T_c^o$ is the transition temperature of 
the noninteracting system. By expanding $F/V$ in powers of $n_o$, we find 
that, to leading order in $na^3$, the shift in transition temperature is 
given by:

\noindent
\begin{equation}
\frac{T_c-T_c^o}{T_c^o} = 1.08 (na^3)^{1/3}.
\label{eq:MF_Correction}
\end{equation}

\noindent
Note that as long as $g>0$, we find that $z<1$ even below the transition 
temperature. Also note that when applying the above analysis to the 
situation discussed in this Paper, e.g. by taking 
$a_{\uparrow\downarrow}=0$, some additional factors of 2 appear that 
lead to replacing the factor 1.08 in Eq. (\ref{eq:MF_Correction}) 
by 0.86, with that equation now relating $(T_c-T_c^o)/T_c^o$ to 
$(na_{\uparrow\uparrow}^3)^{1/3}$.

\noindent {\bf Appendix B:}

In this Appendix we calculate, to leading order the transition temperature 
to the ferromagnetic nonsuperfluid phase. As explained in the main text, 
we treat only the case $g_{\uparrow\downarrow}<0$, and we assume that 
$|g_{\uparrow\downarrow}|$ is large enough that such a phase exists for a 
certain range of temperatures.  We take Eq. (\ref{eq:ferro_criterion}), 
and use the asymptotic behaviour of the functions $g_{3/2}(z)$ and 
$g_{1/2}(z)$ as follows:

\noindent
\begin{equation}
\lim_{z\rightarrow 1^-} \frac{g_{3/2}(z) - g_{3/2}(1)}{\sqrt{1-z}} =
- 3.545,
\end{equation}

\noindent
which gives the asymptotic function:

\noindent
\begin{equation}
g_{3/2}(z) - g_{3/2}(1) \approx - 3.545 \sqrt{1-z}.
\end{equation}

\noindent
The asymptotic limits of the functions $g_{1/2}(z),g_{-1/2}(z),...$ can be
derived using the above approximation for $g_{3/2}(z)$ and the relation
$dg_j(z)/dz=g_{j-1}(z)/z$. Using the asymptotic forms of $g_{3/2}(z)$ and
$g_{1/2}(z)$ in Eq. (\ref{eq:ferro_criterion}), we find that

\noindent
\begin{equation}
\frac{T_{\rm ferro}-T_c^o}{T_c^o} = 1.84 \left( 
n|a_{\uparrow\downarrow}|^3 \right)^{1/3}
\end{equation}

\noindent Comparing the results of Appendices A and B, one would expect 
the boundary between the two regions, i.e. those corresponding to a single 
and double phase transitions, to occur when $|a_{\uparrow\downarrow}| /
a_{\uparrow\uparrow} =0.47$. The value of the ratio 
$|a_{\uparrow\downarrow}| / a_{\uparrow\uparrow}$ for the actual boundary 
will be higher than 0.47 if one takes into account the fact that a 
negative value of $a_{\uparrow\downarrow}$ reduces $T_c$, in agreement 
with the results of our numerical calculation shown in Fig. 1.

\

\noindent {\bf Figures}

Fig. 1: Regions of different behaviour in a gas of spin-1/2 atoms. The $x$ 
and $y$ axes are given by $g_1\equiv g_{\uparrow\uparrow}n/k_BT_c^o$ and 
$g_2\equiv g_{\uparrow\downarrow}n/k_BT_c^o$. In region I we find a single 
phase transition to a superfluid ferromagnetic phase. In region II we find 
two phase transitions, with the ferromagnetic transition occurring at a 
higher temperature than the superfluid transition. Regions III and IV were 
not considered in this Paper because of possible instabilities: the former 
being susceptible to phase separation between the two spin species, and 
the latter susceptible to the whole cloud imploding because the net 
interatomic forces are attractive in that region.

Fig. A1: Schematic plot of the free energy $F(n_o)-F(0)$ of a spinless 
Bose gas as a function of condensate fraction $n_o$ for $T>T_c$ (solid 
line), $T=T_c$ (dashed line), $T_c>T>T_c^o$ (dash-dotted line) $T=T_c^o$ 
(dash-double-dotted line). The dotted line is the $x$ axis and serves as a 
guide to find the temperature of the first-order phase transition. As the 
temperature is lowered, the point that minimizes $F$ jumps discontinuously 
from 0 to a finite value at temperature $T_c$ that is higher than the 
noninteracting transition temperature $T_c^o$.


\begin{thebibliography}{99}
\bibitem{EqualNumbers}  The reason why we are restricting our analysis to 
states where the numbers of atoms in the $\uparrow$ and $\downarrow$ 
states are equal is to simplify the algebra. In the regimes of interest to 
us, the state of the system that minimizes the free energy will satisfy 
that condition even if it is not imposed externally.
\end{thebibliography}
\end{document}